\documentclass[preprint]{aastex}
\usepackage{color}
\usepackage{natbib}
\usepackage{multirow}

\shorttitle{Capturing the heating of type-II spicules with VAULT2.0 and IRIS observations}
\shortauthors{Chintzoglou et al.}

\begin{document}

\title{Bridging the Gap: Capturing the Ly$\alpha$ Counterpart of a Type-II Spicule and its Heating Evolution with VAULT2.0 and IRIS Observations}

%\author{Georgios Chintzoglou\altaffilmark{1}}
%\affil{Lockheed Martin Solar and Astrophysics Laboratory, 
%        3176 Porter Dr, Palo Alto, CA 94304, USA}
%\affil{University Corporation for Atmospheric Research,
%	       Boulder, CO 80307-3000,USA}

%\and

\author{Georgios Chintzoglou\altaffilmark{1,2}    and    Bart De Pontieu\altaffilmark{3}     and     Juan Mart\'{i}nez-Sykora\altaffilmark{4}}
\affil{Lockheed Martin Solar and Astrophysics Laboratory,
	        3176 Porter Dr, Palo Alto, CA 94304, USA}

\and

\author{Tiago M.D. Pereira\altaffilmark{3}}
\affil{Rosseland Centre for Solar Physics, University of Oslo
	        P.O. Box 1029 Blindern, NO-0315 Oslo, Norway}
\and

\author{Angelos Vourlidas\altaffilmark{5}}
\affil{The Johns Hopkins University Applied Physics Laboratory, Laurel, MD 20723, USA}

\and

\author{Samuel Tun Beltran}
\affil{Space Science Division, Naval Research Laboratory, Washington DC 20375, USA}

\altaffiltext{1}{University Corporation for Atmospheric Research, Boulder, CO 80307-3000, USA}
\altaffiltext{2}{\email{gchintzo@lmsal.com}}
\altaffiltext{3}{Institute of Theoretical Astrophysics, University of Oslo, P.O. Box 1029 Blindern, NO-0315 Oslo, Norway}
\altaffiltext{4}{Bay Area Environmental Research Institute, Petaluma, CA 94952, USA}
\altaffiltext{5}{IAASARS, National Observatory of Athens, GR-15236, Penteli, Greece}
\begin{abstract}

	We present results from an observing campaign in support of the VAULT2.0 sounding rocket launch on September 30, 2014. VAULT2.0 is a Ly$\alpha$ (1216\,\AA) spectroheliograph capable of providing spectroheliograms at high cadence. Ly$\alpha$ observations are highly complementary to the IRIS observations of the upper chromosphere and the low transition region (TR) but have previously been unavailable. The VAULT2.0 data provide new constraints on upper-chromospheric conditions for numerical models. The observing campaign was closely coordinated with the IRIS mission. Taking advantage of this simultaneous multi-wavelength coverage of target AR 12172 and by using state-of-the-art radiative-MHD simulations of spicules, we investigate in detail a type-II spicule associated with a fast (300\,km s$^{-1}$) network jet recorded in the campaign observations. Our analysis suggests that spicular material exists suspended high in the atmosphere but in lower temperatures (seen in Ly$\alpha$) until it is heated and becomes visible in TR temperatures as a network jet. The heating begins lower in the spicule and propagates upwards as a rapidly propagating thermal front. The front is then observed as fast, plane-of-the-sky motion typical of a network jet, but contained inside the pre-existing spicule. This work supports that the high speeds reported in network jets should not be taken as real mass upflows but only as apparent speeds of a rapidly propagating heating front along the pre-existing spicule. 
%\bf This abstract could be punched up a bit by adding some statements about how the apparent motions in the Lyman alpha spicule are actually caused by heating fronts, providing support for our recent paper. \rm

\end{abstract}

\section{Introduction}

The chromosphere is one of the least understood layers of the solar atmosphere because of our limited understanding of the governing physical processes. In the last ten to fifteen years, the chromosphere has been propelled to the forefront of solar physics research thanks to spectacular new observations from space with the Solar Optical Telescope (SOT; \citealt{Tsuneta_etal_2008}) onboard the \emph{HINODE} spacecraft, the Interface Region Imaging Spectrograph (\textit{IRIS}; \citealt{dePontieu_etal_2014}), the Very high Angular resolution Ultraviolet Telescope (VAULT; \citealt{Korendyke_etal_2001}), and from the ground with the \textit{Dutch Open Telescope} (DOT; \citealt{Rutten_etal_2004}), the \textit{Interferometric Bidimensional Spectropolarimeter} (IBIS; \citealt{Cavallini_2006}), and the 1-m \emph{Swedish Solar Telescope} (SST; \citealt{Scharmer_etal_2003}), in particular with the CRisp Imaging SpectroPolarimeter (CRISP; \citealt{Scharmer_etal_2008}). The advent of sophisticated numerical simulations, which are beginning to address the complex physics of the optically thick chromospheric plasmas, has helped in the interpretation of the observations (e.g. \citealt{Martinez-Sykora_etal_2017a}). 

A new class of spicules (type-II spicules) has been discovered in Hinode/SOT \ion{Ca}{2} H observations \citep{dePontieu_etal_2007b}. They seem to exhibit much faster speeds as compared to the ``traditional'' class of type-I spicules ($50-100$~km~s$^{-1}$). In \ion{Ca}{2} H filtergrams, type-II spicules appear to be shorter lived (typical lifetimes 10-150\,s) than type-I spicules whereas for higher chromospheric temperatures, e.g., \ion{Mg}{2} h \& k formation temperatures or TR temperatures, e.g., \ion{Si}{4}, the lifetimes are 3-10 minutes \citep{pereira:2014eu}. These spicules are rooted in the chromospheric network and consequently, due to their ubiquity, they are regarded as a significant source of coronal mass and heat transfer \citep{dePontieu_etal_2009}. \citet{Tian_etal_2014} studied counterparts to spicules in the IRIS FUV SJI channels and called them ``network jets''. They measured the plane-of-the-sky speeds with the aid of space-time plots and reported speeds of $\approx$ 80 to 250\,km s$^{-1}$. They concluded that due to these very fast speeds, network jets could be a significant source of mass to the solar wind. On the other hand, \citet{Rouppe_van_der_Voort_2015} using SST and IRIS observations, studied the heating signatures of on-disk type-II spicules and they found that many type-II spicules on the disk are associated with network jets. In addition, they provided Doppler speeds of these on-disk type-II spicules. These speeds (50-75 km s$^{-1}$) seem to disagree with the speeds of \citet{Tian_etal_2014}, despite the similar viewing angles. This discrepancy raises the important question on whether the high speeds reported by \citet{Tian_etal_2014} are apparent speeds and not true mass upflows. \citet{dePontieu_etal_2017} used the state-of-the-art simulations of \citet{Martinez-Sykora_etal_2017a}  and suggested that the speeds of the network jets are likely not caused by real mass motions, but by heating fronts that propagate with those speeds. The network jet speeds would then be understood as apparent speeds (and not real mass motions). This can resolve the discrepancy between the plane-of-the sky motions in images and the line-of-sight velocities from Doppler measurements.

In this paper, we derive new observational constraints from VAULT2.0 and IRIS coordinated observations. The VAULT2.0 rocket experiment (Very high Angular resolution Ultraviolet Telescope 2.0;~\citealt{Vourlidas_etal_2016}) is a Ly$\alpha$ spectroheliograph designed to observe the upper chromospheric region of the solar atmosphere in high spatial ($\sim1\arcsec$) and temporal resolution (8\,s). VAULT2.0 is a reflight of the VAULT rocket and was successfully launched on September 20, 2014 from White Sands Missile Range (WSMR). The VAULT2.0 experiment was complemented by an observing ``campaign'' from ground-based (BBSO, IBIS/NSO, SOLIS) and space-borne (\emph{SDO}/AIA/HMI, \emph{IRIS}, \emph{HINODE}/EIS/SOT) observatories. The importance of the addition of VAULT2.0 Ly$\alpha$ observations is that Ly$\alpha$ is formed at a location in the spectrum where the UV continuum is low, essentially allowing imaging with relatively low photospheric contamination and thus high contrast due to suppressed background emission. This is critical when seeking the on-disk counterparts of type-II spicules. 

This paper is organized as follows: in section 2 we describe the joint VAULT2.0 and IRIS observations of a type-II spicule, in section 3 we discuss the numerical model used to interpret the observations followed by a discussion in section 4. We conclude in section 5.

\section{Campaign Observations}
The VAULT2.0 observing window lasted for about five minutes (18:09 - 18:14 UT). However, the campaign observations were initiated several hours before and concluded a few hours after VAULT2.0 flight. 
During its flight, VAULT2.0 observed a failed eruption and captured cooling downflows and bright ribbons around the target AR12172 (\citealt{Chintzoglou_etal_2017a}). At the time of the flight the AR was at heliographic coordinates S15$\degr$W55$\degr$. The VAULT2.0 large (384$\arcsec\times$256$\arcsec$) Field of View (FOV), captured substantial areas of quiet Sun (QS) around the target AR. The spatial resolution was determined post-flight ($\lesssim1\farcs2$; \citealt{Vourlidas_etal_2016}) and 34 images were recorded at an 8\,s cadence.

\begin{figure*}
\epsscale{1.0}
	\plotone{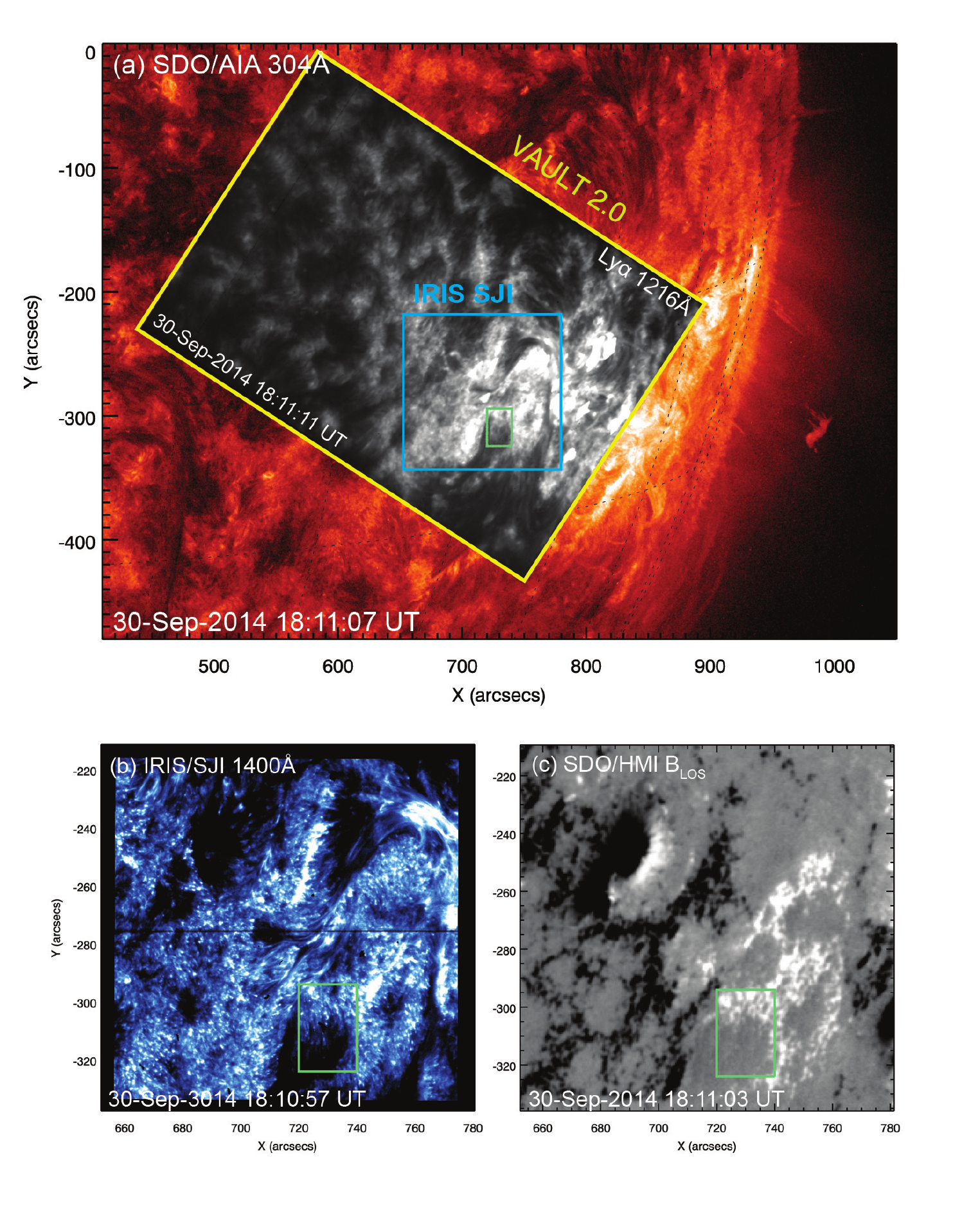}%{Fig1.eps}
	\caption{(a) Context image from AIA/SDO at 304\,\AA\ showing the FOV of VAULT2.0 and IRIS. The green box shows a region of 20$\arcsec\times$30$\arcsec$ centered at the location of the Type-II spicule co-observed by VAULT2.0 and IRIS. (b) The FOV of IRIS SJI in 1400\,\AA. (c) Line of sight magnetogram from HMI/\textit{SDO} for the FOV of panel (b) saturated at $\pm$500\,G. The location where the type-II spicule and network jet occurs (green box) is adjacent to a positive polarity plage region.}
\label{CONTEXTFIG}
\end{figure*}

IRIS supported the observational campaign by providing fast time-cadence (13\,sec) image series from its slit-jaw imagers (in short, SJI). IRIS was in a ``sparse'' ($1\arcsec$ separation) 8-step raster mode (OBS-ID 3820255277), centered at (x,y)=(711$\arcsec$,\,-281$\arcsec$) and observed the target between 17:13 - 20:04 UT. The SJI passbands covered the lower TR at 1400\,\AA\ (dominated by \ion{Si}{4} ion emission) and at 1330\,\AA\ (dominated by \ion{C}{2} ion emission) with mean formation temperatures of $\approx$ 80,000\,K and $\approx$ 30,000\,K respectively (see \citet{Rathore_etal_2015} for further discussion on the dominant temperatures of plasma emission in \ion{C}{2} 1330\,\AA). The FOV of these IRIS SJI images (120$\arcsec\times$120$\arcsec$) is smaller than that of VAULT2.0, but large enough to contain the core of the target AR as well as some QS region north and south of the AR. The SJI pixel size is $0\farcs167$ pix$^{-1}$ and the data were corrected for flat-field, dark current and geometric distortions as described in \citet{dePontieu_etal_2014}. In Figure \ref{CONTEXTFIG} we show the FOVs of each instrument superimposed on a 304\,\AA\ context image from the \textit{Atmospheric Imaging Assembly} (\textit{AIA}; \citealt{Lemen_etal_2011}) onboard the \textit{Solar Dynamics Observatory} (\textit{SDO}; \citealt{Pesnell_etal_2012}). Qualitatively the Ly$\alpha$ observations show similarities in structures observed with the 304\,\AA\ passband of SDO/AIA and also with the SJI 1400\,\AA\ and 1330\,\AA\ passbands of IRIS (compare panels a and b of Figure~\ref{CONTEXTFIG}). In the QS the chromospheric network is the dominant structure while in the location of the AR an inverse U-shape filament is seen as a dark feature, surrounded by emission from the nearby plage.

%% DESCRIBE THE OBSERVATIONS FIGURE 2

\begin{figure*}
\epsscale{1}
	\plotone{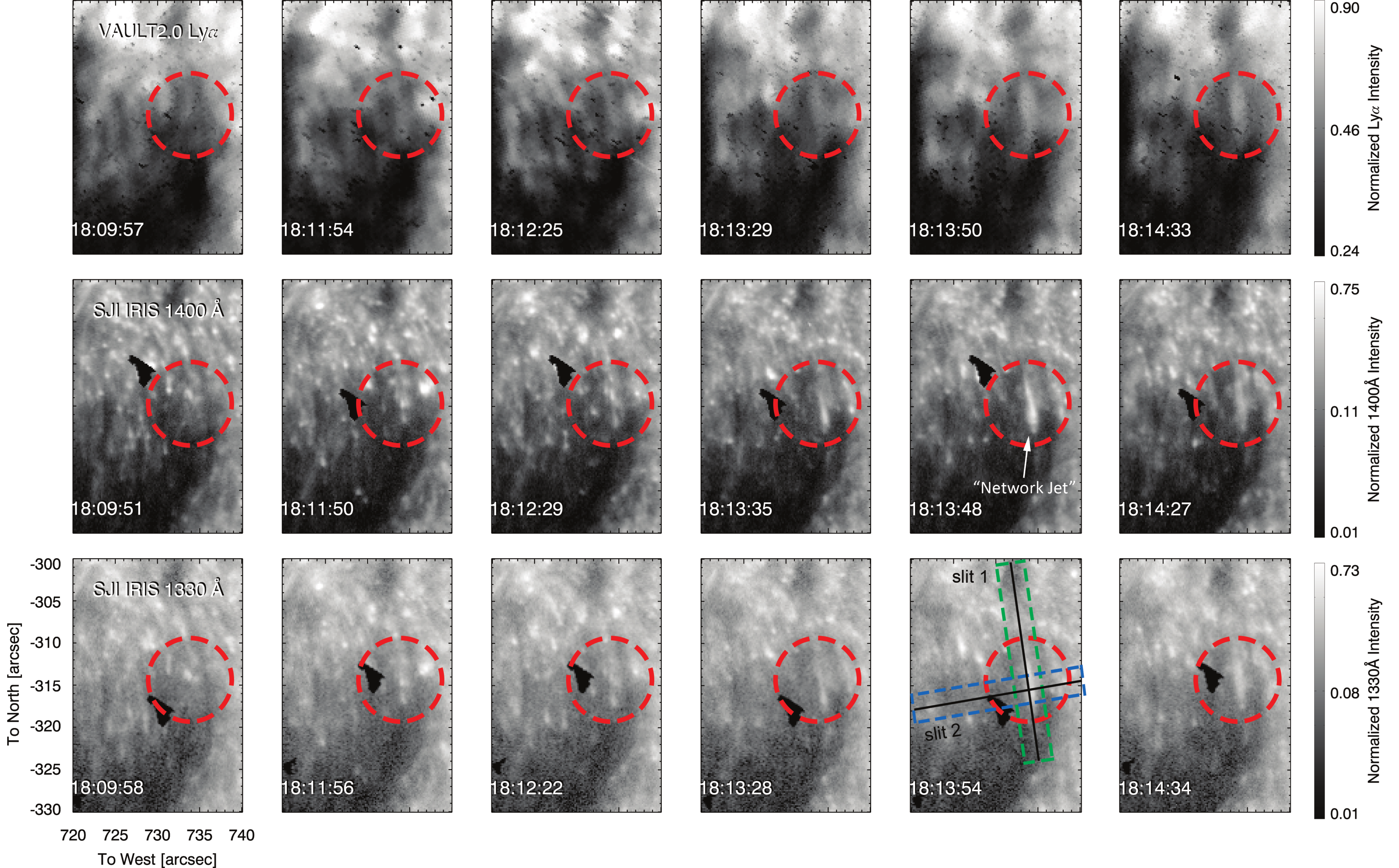}%{VAULT_IRIS_multipanel_new.eps}
  \caption{
	  The time-evolution of a network jet observed in Ly$\alpha$ 1216\,\AA\ by VAULT2.0 (top row) and with the IRIS slit-jaw imager in \ion{Si}{4} 1400\,\AA\ (middle row) and \ion{C}{2} 1330\,\AA\ (bottom row). The FOV corresponds to the green box of Figure~\ref{CONTEXTFIG}. Intensities for each passband $I$ are normalized as $I/I_{max}$ with $I_{max}$ the strongest intensity in the FOV of the entire time series. Scaling is logarithmic (base 10) for all panels. The sharp dark features in the IRIS FOVs are particles on the slit-jaw detector. The network jet occurs at the time shown in the last columns of \ion{Si}{4} 1400\,\AA\ and \ion{C}{2} 1330\,\AA . The Ly$\alpha$ counterpart of the network jet is shown inside the red dashed circle in the top row and it seems to pre-exist (at least in panel at 18:11:54 UT) before the appearance of the network jet (compare content in red dashed circles between wavelengths). The green dashed rectangle shown in panel 18:13:54 marks the area containing the network jet and the blue dashed rectangle marks the area intersecting the network jet at a 90$\degr$ angle. Pixels along the dark lines inside these locations are used in the space-time plots of Figure~\ref{Obs_Slitstack}. 
  }
 \label{VAULT_IRIS_OBS}
\end{figure*}

Since the campaign includes only on-disk observations, we restricted our efforts in finding type-II spicules on-disk. This restriction, combined with the relatively small IRIS SJI FOV, made for a challenging analysis since spicules are best seen against low background emission (which is best for spicules at the limb). After carefully inspecting the image series in the common FOV of VAULT2.0 and IRIS for spicules, we identified a fast network jet in the IRIS 1400\,\AA\ and 1330\,\AA\ passbands in the south of AR12172 (green box in Figure~\ref{CONTEXTFIG}). In Figure~\ref{VAULT_IRIS_OBS} we show three snapshots of the area around the network jet. The $20\arcsec\times$30$\arcsec$ FOV contains chromospheric material with both dark and bright features (see Figure~\ref{CONTEXTFIG}). These features form a darker background against which a sudden brightening, typical of a network jet, can be identified on-disk. The network jet is shown in the bottom right panel of Figure~\ref{VAULT_IRIS_OBS} (indicated with a white arrow). Based on the SJI images, the width of the jet is $\sim 0\farcs5$.

\begin{figure*}
\epsscale{0.8}
	\plotone{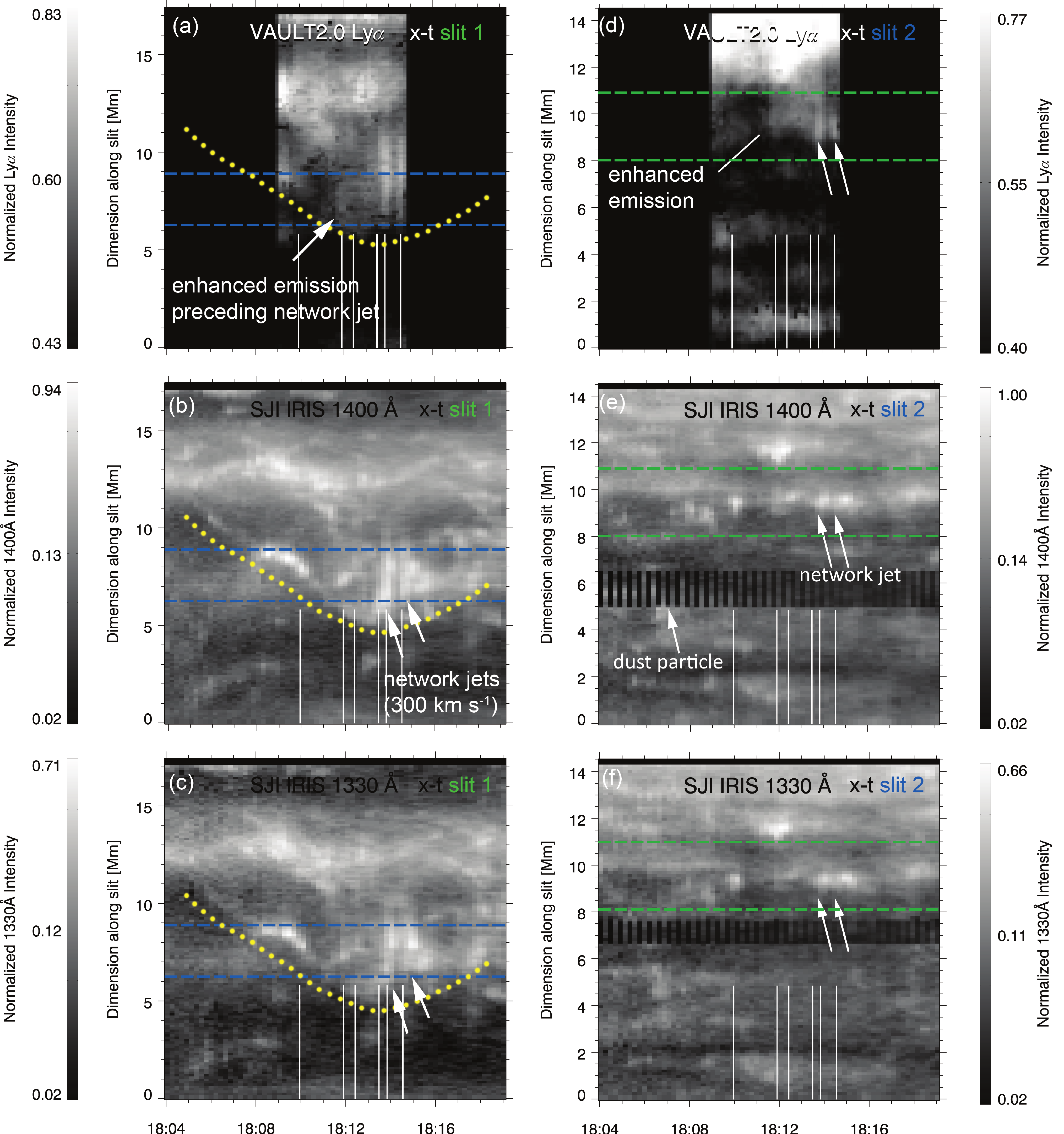}%{fig3.eps}%{VAULT_IRIS_slitstack_new_2.eps}
  \caption{
	  Space-time plots composed for all wavelengths for slits 1 and 2 shown in Figure~\ref{VAULT_IRIS_OBS} (panel at 18:13:54). Left panels (a, b, c) show the evolution of the brightenings along slit 1 for each passband and the right panels show the evolution along slit 2, i.e. at a cross-section with the network jet. To guide the eye with reference to Figure~\ref{VAULT_IRIS_OBS}, we show the intersection of the rectangular area of slit 2 with slit 1 as blue dashed lines. Similarly, in the right panels we show with green dashed lines the area around slit 1 as intersected by slit 2 (containing the cross section of the network jet once it occurs). Intensity scaling is logarithmic (base 10). Times when VAULT2.0 was not observing are shown with black color. The bottom part of the spicule is at the top of slit. In the 1400\,\AA\ and 1330\,\AA\ space-time plots we see that the type II spicule manifests itself in the form of a parabolic trace (yellow dotted curve superposed in all left panels for reference). The spicule seems to undergo a recurring second brightening along its length 40\,s after the first brightening, also shown in 1216\,\AA . The selected times of the panels in Figure~\ref{VAULT_IRIS_OBS} are shown with white vertical lines. The type-II spicule in Ly$\alpha$ shows emission preceding the sudden brightening/network jets seen in IRIS by almost two minutes (panel a; arrow pointing to location above parabolic trace).
	  }
  \label{Obs_Slitstack}
\end{figure*}

The space-time plot of Figure~\ref{Obs_Slitstack} shows a parabolic profile (visible in both IRIS passbands but best seen in \ion{Si}{4} 1400\,\AA\ due to its better signal-to-noise ratio). This is compatible with the presence of a type II spicule, with the bright TR emission outlining the spatial extent of the spicular material, as also shown in~\citet{dePontieu_etal_2017}. This may be associated with a heating front at TR temperatures as it travels along the spicule's axis~\citep{Martinez-Sykora_etal_2017a}. 
When the spicule is fully evolved ($\sim$ 40\,s later), it brightens along its whole length (18:13:08 UT) in the 1400\,\AA\ and 1330\,\AA\ SJI images. Emission re-occurs about a minute after the first peak of emission, at 18:14:40 UT. This resurgence leaves two almost vertical streaks in the space-time plot of 1400\,\AA\ and 1330\,\AA\ in Figure~\ref{Obs_Slitstack}. Their vertical extent corresponds to the length of the spicule, $\sim$ 4\,Mm projected on the plane of the sky. From these space time plots we deduce an apparent speed of $\sim 300$ km~s$^{-1}$. 

The VAULT2.0 observations reveal a slighlty different behavior than the IRIS observations. Ly$\alpha$ emission peaks simultaneously with the \ion{Si}{4} 1400\,\AA\ and \ion{C}{2} 1330\,\AA\ emission as shown in the space-time plots of Figure~\ref{Obs_Slitstack}. However, while the apparent parabolic profiles in the IRIS passbands develop, there is sudden Ly$\alpha$ emission over the full length of the spicule for up to 120\,s before the first brightening in 1400\,\AA\ and 1330\,\AA\ at 18:13:48 UT (emission between the 18:12 and 18:14 vertical bar, top left panel in Figure~\ref{Obs_Slitstack}). The feature can be seen in the VAULT2.0 frame in the middle top panel of Figure~\ref{VAULT_IRIS_OBS} (red dashed circle). In addition, the Ly$\alpha$ structure has the same linear extent as the IRIS-observed type-II spicule, which suggests that material appears to pre-exist at the location where the network jet occurs, but at temperatures lower than the TR temperatures observed by IRIS.

The measured plane-of-the-sky speed of $\sim 300$ km~s$^{-1}$ lies on the high-end of the speeds of network jets reported by \citet{Tian_etal_2014} and \citet{Narang_etal_2016}. In this particular case, the brigthening is happening in a spicule in an active region. In the following section, we are going to investigate this case with an advanced radiative MHD model. 

\section{Bifrost Model of a Spicule}

To address the questions raised by our analysis we compare the observations to simulated spicules \citep{Martinez-Sykora_etal_2017a}. The Bifrost code \citep{Gudiksen:2011qy} solves the MHD equations while including the radiative losses \citep{Hayek:2010ac,carlsson:2012uq}, thermal conduction along the magnetic field and the ion-neutral interaction effects by adding the ambipolar diffusion and the Hall term in the induction equation (\citealt{Martinez-Sykora:2012uq,Martinez_Sykora_etal_2017gol} for details on the ambipolar diffusion on these models). This model solves the single-fluid MHD equations and takes into account ion-neutral interaction effects. In short, in a partially ionized magnetized gas, the ambipolar diffusion describes the decoupling of the bulk of the fluid with magnetic flux. This is a consequence of neutrals being able to move ``freely'' while ions are coupled to the magnetic field. Due to collisions, ions can also be decoupled from the magnetic field. Note that the ambipolar diffusion allows to diffuse the magnetic field and dissipates magnetic energy into thermal energy. This numerical model is capable of self-consistently producing type-II spicules \citep{Martinez-Sykora_etal_2017a}. 

The simulation analyzed here is the 2.5D radiative MHD model described in detail in \citet{Martinez-Sykora_etal_2016, Martinez-Sykora_etal_2017a}. The numerical domain ranges from the upper layers of the convection zone ($3$~Mm below the photosphere) to the self-consistently maintained hot corona ($40$~Mm above the photosphere) and $90$~Mm along the horizontal axis. The spatial resolution is 14\,km along the horizontal axis. For the vertical direction, the grid is non-uniform allowing a finer grid size in locations where it is necessary to resolve features, such as in the photosphere to the TR, and coarser in the corona and deeper into the convection zone (i.e. variable grid size ranging from 12$-$50\,km). This simulation shows the formation of very fast network jets seen in synthesized \ion{Si}{4} emission~\citep{dePontieu_etal_2017}. In addition, we derive synthetic Ly$\alpha$ observables to interpret the thermal evolution from the observations.

For this work we calculated the synthetic \ion{Si}{4} intensity assuming
ionization equilibrium and optically thin approximation as in \citet{Hansteen:2010uq}. Ly$\alpha$ profiles were calculated in detail by solving
the full non-LTE radiative transfer problem on a column by column basis
with the RH 1.5D code \citep{Uitenbroek:2001dq,Pereira:2015th}. A
five-level plus continuum model hydrogen atom was used. The 1.5D
approximation was used to keep the computational costs down. To compare
with VAULT2.0 observations, we integrated the synthetic Ly$\alpha$ line
profiles over a Gaussian bandpass with 80\AA\ FWHM. Because of this
integration, the mean intensity is more important than the spectral line
shapes. While Ly$\alpha$ suffers from partial redistribution (PRD)
effects \citep{Milkey_Mihalas_1973}, testing a single snapshot we find
that when using the 1.5D approximation, calculations assuming complete
redistribution (CRD) give a mean intensity that is closer to the full 3D
PRD approach. The reason is that PRD reduces the intensity in the line
wings, while 3D effects do the opposite \citep{Sukhorukov_Leenaarts_2017}.
Therefore, our adopted 1.5D CRD approach gives a reasonable
approximation when studying wavelength-integrated Ly$\alpha$ profiles.

\begin{figure*}
\epsscale{1}
	\plotone{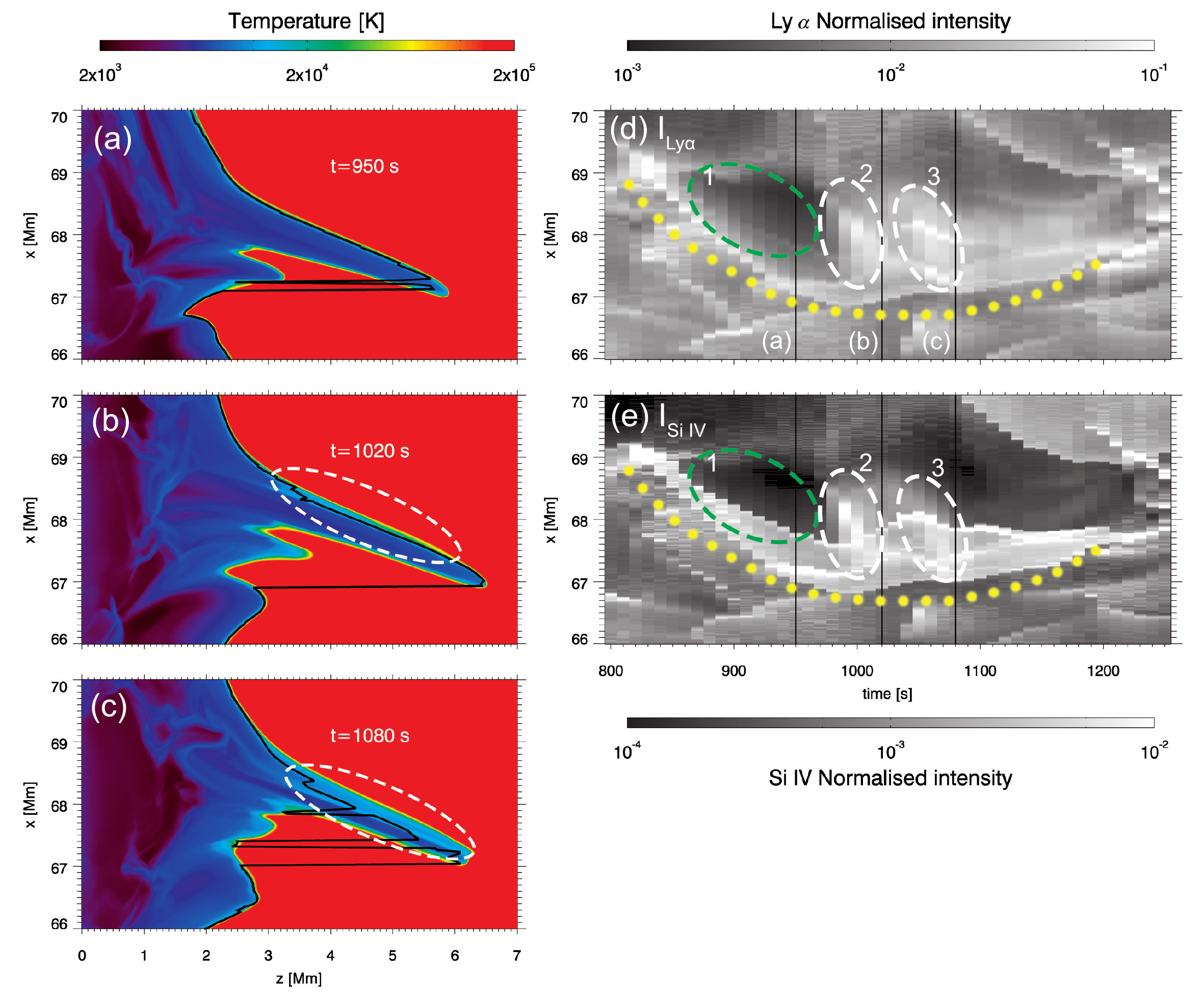}%{fig4.eps}%{ly_si_65_arrows.pdf}
	\caption{Panels a-c: Temperature maps for a simulated type-II spicule at three time instances (shown with black lines in panels d,e). The horizontal dimension, x, is placed in the vertical axis to account for a spicule that grows southward (as in the observations). The photospheric surface is at z=0. The black curve shows the height of $\tau$=1 for the peak intensity of Ly$\alpha$ for an observer viewing from the top-down towards the solar surface (in this arrangement of the maps, the observer's view is from right to left). While spicular material is already suspended up to 6\,Mm from the photosphere, Ly$\alpha$ emission comes from different parts along the spicule as it heats up. Panels d-e: Ly$\alpha$ and \ion{Si}{4} space-time plots of their normalized emergent intensities, $I/I_{max}$ with $I_{max}$ the strongest intensity in the entire time series. The spatial axis is the vertical axis in a-c panels. The intensity scaling in all panels is logarithmic (base 10). Note the parabolic trace left in the space time plots due to the evolving spicule (yellow dotted curve). Note the higher normalized intensity of Ly$\alpha$ as compared to \ion{Si}{4} (inside the green dashed oval 1; before the first network jet). The occurence of the network jets is marked with white dashed ovals 2 and 3 (see text). 
}\label{Bifrost_stack}
\end{figure*}

In Figure~\ref{Bifrost_stack} we present temperature maps from the simulation at three consecutive instances of the type-II spicule (panels a, b, and c) and synthesized Ly$\alpha$ and \ion{Si}{4} emergent intensities in space-time plots (as observed from the top of the box). Similarly to the observations in \ion{Si}{4}, a bright point-like front travels with the growing spicule as seen in the space-time plots (forming a parabolic profile, yellow dotted envelope in panels d and f). This is arises from a moving front of TR emission at the top of the spicule (panel a), enhancing the emissivity in both simulated passbands (Ly$\alpha$ and \ion{Si}{4}). However, up to that time, in the region enclosed by the yellow-dotted parabolic envelope (inside green dashed oval 1 in panels d and e), there seems to be an order of magnitude difference in the normalized intensities between Ly$\alpha$ and \ion{Si}{4}, with Ly$\alpha$ emission being higher than \ion{Si}{4}. This enhanced emissivity of Ly$\alpha$ comes from the spicular column (above $\tau = 1$), similarly to the enhanced emission seen in the observed Ly$\alpha$ space time profile before the occurrence of the network jet in \ion{Si}{4} (panel a and b in Figure~\ref{Obs_Slitstack}). Then, at $\sim$990\,s a sudden brightening occurs along the whole spicule in \ion{Si}{4} (and Ly$\alpha$) caused by a sudden temperature increase (up to $\sim10^4$~K) in a very narrow region on the right hand side of the spicule (panel b). This corresponds to the bright linear feature inside the dashed oval 2 in the space time plots in panels d and e.

In the temperature panels, the formation height of Ly$\alpha$ ($\tau = 1$, shown with black solid line in panels a-c) is exactly at the location of the region which undergoes the temperature increase (dashed oval, panel b). 
Around 70\,s later (t=1080\,s; dashed oval 3 in space-time plots in panels d and e) a second emission enhancement occurs associated with an increase of temperature to $\sim10^4$~K of most of the spicular plasma (panel c, dashed oval).

%\bf DO WE GET EMISSION IN LYALPHA BEFORE SIIV IN SIMULATIONS? \rm

\begin{figure*}
\epsscale{1.0}
	\plotone{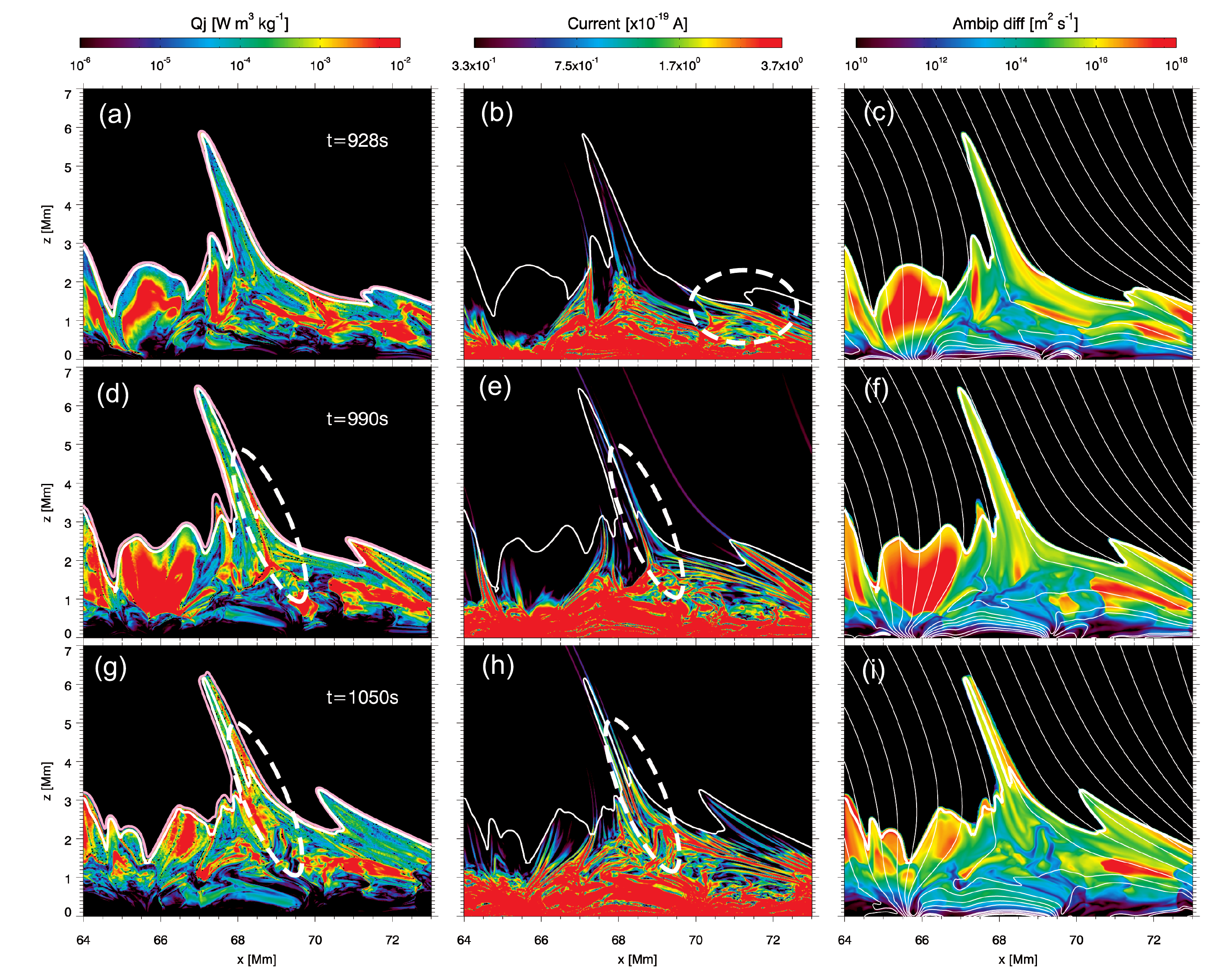}%{fig5.eps}%{qj_65_9p_arrows.pdf}
\caption{
	The Joule heating, $Q_j$, the magnitude of electrical current, $|\mathbf{J}|$, and ambipolar diffusion for the Type-II spicule of Figure~\ref{Bifrost_stack} arranged in columns at three time instances of the simulation (rows at $t=$ 928\,s, 990\,s, 1050\,s). The thick white (all panels) and pink (left) lines show $T=10^4$~K and $T=10^5$~K isocontours, respectively. Selected magnetic field lines are shown in the right panels (thin white lines). Note the evolution of ambipolar diffusion and the electrical currents with the receding temperature isocontour over time. At t=928\,s electric currents between $x=$ 71 $-$ 73\,Mm intensify (panel b; white dashed oval) and at $t=$ 990\,s extend radially outwards along the spicule (dashed oval in panel e) increasing the Joule heating and subsequently the temperature of the outer part of the spicule (dashed ovals in panels d,g). 
}\label{Spicule_heating}
\end{figure*}

To further investigate the physical processes responsible for the simulated event, we explored the Joule Heating (left column, Figure~\ref{Spicule_heating}), electrical current ($|\mathbf{J}|$, middle), and ambipolar diffusion (right) at selected times before the times of the panels a, b, and c of Figure~\ref{Bifrost_stack}. 
An electrical current is traveling within the chromosphere (directed from right to left as seen in Figure~\ref{Bifrost_stack}) at Alfv\'{e}nic speeds. At $t=928$~s the current front is at $x\approx70.8$~Mm (shown with a dashed oval in panel b) and at $t=990$~s has spread out along the very right hand side of the spicule (inside oval in panel e). Due to ambipolar diffusion (Figure~\ref{Spicule_heating}, right column) the current is dissipated rapidly and heats the plasma (left panels) in a very localized region along the spicule. The heating time scale of a few seconds is a result of the rapid propagation and of the rapid dissipation of the currents, i.e., at an Alfv\'{e}n speed between 300-500~km~s$^{-1}$. Following this Alfv\'{e}nic wave, a wider current front is covering most of the spicule at $t=1050$~s.

\section{Discussion}
%% Putting everything together -- 

The existence of material in Ly$\alpha$ temperatures before the manifestation of network jets challenges the idea that the plane-of-the-sky motions of network jets are real mass motions of spicules seen in the TR. Previously, \citet{Tian_etal_2014} and \citet{Narang_etal_2016} performed a statistical study of a large sample of network jets seen in IRIS \ion{C}{2} by measuring the plane-of-the-sky speeds and they found a distribution of speeds between 80-300 km s$^{-1}$. \citet{Rouppe_van_der_Voort_2015}, conducted an analysis of the heating signatures of on-disk type-II spicules using H$\alpha$ data from the SST with \ion{Si}{4} and \ion{C}{2} from IRIS. In their work they reported Doppler velocities much lower than those in \citet{Tian_etal_2014}, i.e. 50-75 km s$^{-1}$. We should stress here that both studies dealt with on-disk case of network jets and type-II spicules, with similar viewing angles. The difference in the reported speeds between the two studies is striking. One would expect a close match of speeds deduced by the two methods (i.e. plane-of-the-sky speeds and Doppler-deduced speeds) if those motions were real mass motions. If there are real mass motions of $\sim150-300$\,km s$^{-1}$ then, given that network jets are observed to be roughly radially oriented, then the doppler speeds should reveal these speeds or at least show an order-of-magnitude agreement. The observations from VAULT2.0 and IRIS suggest that type-II spicules pre-exist at temperatures visible in Ly$\alpha$, and that some of the plasma in these type-II spicules suddenly gets heated to higher, TR temperatures, which is revealed as rapidly propagating linear features in the TR passbands observed by IRIS (i.e. network jets). That is, our analysis does not support the interpretation that the speeds are real mass motions but rather suggests that they arise from thermal fronts, which heat the spicular plasma. 

%The parabolic profiles seem to suggest that the slow upward motion of the spicular system occurs at speeds $\approx$ 15-20 km s$^{-1}$, which considering the viewing geometry (50$\degr$ from disk center) is more conformant to the magnitude of the speeds deduced by the doppler effect (as in \citealt{Rouppe_van_der_Voort_2015}).  

%The succession of events in the simulation shows striking resemblance to the VAULT2.0 observations. Material is already suspended from the surface up to several Mm and heating governs the phenomenology as described in the observations in the following way: the fast thermal evolution within the spicules is due to rapid dissipation via ambipolar diffusion of transverse waves and electric currents traveling along the spicule at Alfv\'{e}nic speed. This conclusion provides further support for the numerical simulations. 

Furthermore, we conclude that since type-II spicules seem to have much lower mass flow speeds than the apparent speeds of network jets (which are not mass flows), this may also have implications for estimates of the contributions of spicules to the solar wind mass flux. As discussed in \citet{Tian_etal_2014}, using a mean jet speed from their sample of 150 km s$^{-1}$ they estimate a spicular contribution to the solar wind of (2.8 - 36.4)$\times$10$^{12}$ g s$^{-1}$ if all mass of the network jets contributes to the solar wind. In addition, they acknowledge that the lack of coronal observations of sufficient sensitivity makes it difficult to assess whether the aforementioned rate is realistic contribution to the solar wind. They concluded that the above mass contribution rate is $\approx$ 2 to 24 times larger than the total mass loss associated with the solar wind (ranging from 1.2 - 1.9 $\times$ 10$^{12}$ between solar minimum and solar maximum respectively). Using the same assumptions here, we would argue that by using the conservative Doppler speeds of about 50 km s$^{-1}$ (as measured in the analysis of \citet{Rouppe_van_der_Voort_2015} on type-II spicules) as to be a more realistic measure for the mass upflows in type-II spicules (instead of the apparent speeds of network jets), this would reduce the values to about a third, i.e. $\approx$ 0.6 - 8 times the total loss rate of the solar wind. However, neither the \citet{Tian_etal_2014} values nor these values may be an accurate measure of the mass contributing to the solar wind, since it is not known what fraction of the spicular mass is heated to million degree temperatures. We need to stress here that addressing the question of spicular contribution to the solar wind in the way of \citet{Tian_etal_2014} is complicated by the fact that, just like with coronal heating, the contribution of the spicules may not be limited to just providing mass, but could also include generation and propagation of currents, which can then dissipate in the wind and lead to extra evaporation. 
Indeed, parabolic profiles usually suggest that the suspended mass comes down. Our observations, with unique temperature coverage and spatiotemporal resolution, support the view presented in \citet{dePontieu_etal_2017}, i.e. that network jets are primarily apparent plane-of-the-sky motions arising from enhanced emission of rapidly propagating heating fronts along pre-existing spicules, rather than actual mass motion upwards.

\section{Conclusion}

We observed the occurrence of a type-II spicule in the common FOVs of VAULT2.0 and IRIS SJI imagers. The spicule contains Ly$\alpha$ material at lower intensities for about two minutes prior to its counterpart in \ion{Si}{4} and \ion{C}{2} (Figures~\ref{VAULT_IRIS_OBS} and \ref{Obs_Slitstack}). Once the type-II spicule reaches TR temperatures, it brightens twice within 52\,s in all three passbands (Figure~\ref{Obs_Slitstack}).

We employed a state-of-the-art 2.5D MHD Bifrost simulation of a type-II spicule (\citealt{Martinez-Sykora_etal_2017a}). According to this model, ambipolar diffusion facilitates the dissipation of electrical currents whithin the type-II spicules causing the temperature to increase rapidly. The synthesis of observables from this simulation shows precedence of significant emission for Ly$\alpha$ as compared to \ion{Si}{4} emission in the spicule. The rapid temperature changes have an impact on the emissivity of the different ions. Two brightenings in the same spicule arise from two episodes of current dissipation within the spicule. 

In conclusion, the precedence of Ly$\alpha$ emission two minutes before the brightening in \ion{Si}{4} 1400\,\AA\ and \ion{C}{2} 1330\,\AA\ points to the existence of a ``Ly$\alpha$ counterpart'' for the network jets. Our unique observations elucidate the heating process and the morphology of the type-II spicules. \citet{dePontieu_etal_2017} have provided evidence that the speeds of network jets reported by \citet{Tian_etal_2014} and \citet{Narang_etal_2016} are not true jet speeds and conversely not mass upflows at those speeds. Our analysis of VAULT2.0 observations provides additional evidence in agreement with \citet{dePontieu_etal_2017} and suggests that spicular material pre-exists at chromospheric temperatures as a Ly$\alpha$ spicule. That spicule heats up to TR temperatures due to rapidly moving heating front which results in the apparent (and not real mass motion) speeds in the form of fast network jets within the same spicule.

\acknowledgments
We thank the anonymous referee for careful reading and comments that improved the article. We gratefully acknowledge support by NASA grants NNM12AB40P, NNX16AG90G, and NNG09FA40C (IRIS). IRIS is a NASA Small Explorer mission developed and operated by LMSAL. The simulations have been run on clusters from the Notur project, and the Pleiades cluster through the computing project s1061, s1472 and s1630 from the High End Computing (HEC) division of NASA. This research was supported by the Research Council of Norway through its Centres of Excellence scheme, project number 262622, though grant 170935/V30, and through grants of computing time from the Programme for Supercomputing.We also acknowledge support form NSF grant AST 1714955 and NASA GI grant NNX17AD33G. The VAULT2.0 project and launch operations (G.C., A.V., S.T.B.) were supported by NASA NNG12WF67I. The participation of A.V in this work was partially funded by NRL grant N00173-16-1-G029.

%\bibliography{../references}%%FOR MAC ``/home'' -> ``/Users'
%\bibliographystyle{aa}
%\bibliography{../aamnemonic,../collectionbib,../references}%%FOR MAC ``/home'' -> ``/Users'

\end{document}